\documentclass[a4paper,noarxiv,unpublished,twocolumn]{quantumarticle}
\pdfoutput=1

\usepackage[utf8]{inputenc}
\usepackage[english]{babel}
\usepackage[T1]{fontenc}
\usepackage{amsmath}
\usepackage{hyperref}
\usepackage{tikz}
\usepackage{lipsum}

\usepackage[numbers,sort&compress]{natbib}

\usepackage{mathtools}
\usepackage{amssymb}
\usepackage{braket}
\usepackage{amsfonts}

\usepackage{qcircuit}
\usepackage[linesnumbered,ruled,vlined]{algorithm2e}
\usepackage{textcomp}
\usepackage{wasysym}
\usepackage{graphicx}
\usepackage{caption}
\usepackage{subcaption}

\usepackage{amsthm}

\newtheorem*{theorem-non1}{Quantum Amplitude Interpolation}
\newtheorem*{theorem-non2}{Controlled Weighted Sum}
\newtheorem*{application1}{Expected Value}

\emergencystretch=\maxdimen
\graphicspath{ {./images/} }

\renewcommand{\floatpagefraction}{0.7}

\begin{document}

\title{Quantum Amplitude Interpolation}

\author{Charlee Stefanski}
\orcid{0000-0001-9856-5955}
\affiliation{Wells Fargo}
\affiliation{UC Berkeley}
\author{Vanio Markov}
\affiliation{Wells Fargo}
\author{Constantin Gonciulea}
\orcid{0000-0001-5870-4586}
\affiliation{Wells Fargo}

\maketitle
\begin{abstract}
    In this paper we present a method for representing continuous signals with high precision by interpolating
    quantum state amplitudes. The method is inspired by the Nyquist-Shannon sampling theorem, which links continuous
    and discrete time signals.

    As an application, for positive integers $n$ and $m$, and a given discrete function with real values
    $f: \{0, \mathellipsis, 2^n-1\} \rightarrow [-2^{m-1}, 2^{m-1})$, this method enables the estimation of weighted sums of
    hashed function values $\sum_{k=0}^{2^n-1} w_k h(f(k))$, with $w_k \in \mathbb{R}$ for $0 \le k < 2^n$ and $h:
    [-2^{m-1}, 2^{m-1}) \rightarrow \mathbb{R}$.

    This method extends our previous method of computing generalized inner products from integer-valued functions
    to real-valued functions.

\end{abstract}
\maketitle

\section{\label{sec:introduction}Introduction}

The state of a quantum system or register with $n$ qubits is represented by a set of complex numbers, called amplitudes,
one for each of the $2^n$ possible outcomes of a system measurement. An outcome corresponds to an integer between
$0$ and $2^n-1$ whose binary representation is derived from the individual qubit measurements of $0$ or $1$.

Mathematically, the state is a function $f: \{0, \mathellipsis, 2^n-1\} \rightarrow \mathbb{C}$
with the property $\sum_{k=0}^{2^n-1} |f(k)|^2 = 1$,
typically expressed using the ket notation for the computational basis:

\begin{equation*}
    \sum_{k=0}^{2^n-1} f(k) \ket{k}_n.
\end{equation*}

For a given integer $j \in \{0, \mathellipsis, 2^n-1\}$ the computational basis state $\ket{j}_n$
can be a convenient digital encoding of $j$ into quantum computations when appropriate.
A common strategy for dealing with non-integers is to increase the number of qubits to a level that allows
the approximation of a real value by its closest integer.
This is the strategy used in the Quantum Phase Estimation algorithm.

In this paper we introduce a method for interpolating quantum amplitudes that enables:
\begin{enumerate}
    \item Exact results of certain quantum computations involving real numbers by interpolating their discretized
    versions.
    \item Reducing the number of qubits needed to approximate non-integer values digitally encoded in the state of a
    quantum system.
    \item Reducing the error of quantum computations that involve real numbers compared to direct discretization
    methods.
\end{enumerate}

\bigskip
The paper is organized as follows:

Section~\ref{sec:prelim} introduces notation and concepts used throughout the paper, as well as an overview of
Classical Intepolation theory.

Section~\ref{sec:methods} presents methods for quantum interpolation, phase corrected real-value number encoding, and
phase corrected real-valued discrete function encoding. This section also includes an extension of the generalized
inner product method introduced in ~\cite{GenInnerProduct} which uses quantum interpolation to compute the weighted
sum of real-valued functions.

Section~\ref{sec:applications} presents applications of the quantum interpolation method.

Section~\ref{sec:experiments} presents results of running some applications on IBM Q
hardware~\cite{IBMQServices}.

Section~\ref{sec:conclusions} contains concluding remarks and considerations for future work.

\section{\label{sec:prelim}Preliminaries}

\subsection{\label{subsec:classical_interpolation}Classical Interpolation Theorem}

To represent a continuous function $f: \left[ 0, T \right] \rightarrow\mathbb{R}$ where $T > 0$ in an $n$-bit
digital computer we can uniformly sample $N=2^n$ values of $f$ into a vector $\textbf{x}=(x_0, \mathellipsis,
x_{N-1})$ where $x_k = f(k\frac{T}{N})$ for $k \in \{ 0, \mathellipsis , N-1 \}$. This approach introduces a
discretization error up to $\max\limits_{1 \le k < N}|f(k \frac{T}{N})-f((k-1) \frac{T}{N})|$. Under specific
conditions the vector \textbf{x} will carry enough information in order to define the function values between the
sampling points ~\cite{uiocourse, kammler_2008}.

Let us assume that the Fourier expansion of $f$ is finite:

\begin{equation*}
    f(t) = \sum_{l=-L}^{L} z_l e^{i 2\pi l\frac{t}{T}}
\end{equation*}

for $t \in [0, T]$ where $z_l$ are Fourier coefficients~\cite{BoggessSamplingChapter}.

Such functions $f$, with a finite frequency spectrum are called \textit{band-limited} ~\cite{BoggessSamplingChapter}.
Assume the vector $\textbf{x}=(x_0,\mathellipsis,x_{N-1})$ contains $N=2L+1$ uniform samples from the interval $[0,
T]$ and $\textbf{y}=(y_0,\mathellipsis,y_{N-1})$ is the DFT (Discrete Fourier Transform) of $\textbf{x}$:

\begin{equation*}
    y_j = \frac{1}{\sqrt{N}} \sum_{k=0}^{N-1} x_k e^{-i 2\pi j \frac{k}{N}}
\end{equation*}

for integers $0 \le j < N$.

It can be proven ~\cite{uiocourse, kammler_2008} that $z_l = \frac{1}{\sqrt{N}} y_{l+L}$ for $-L \le l \le L$.
Using this fact, a periodic, band-limited function $f$ can be reconstructed at any non-integer $t \in [0, T)$ from $N
\ge 2L+1$ samples as follows:

\begin{equation}
    \label{eqn:classical_interpolation}
    f(t) = \frac{1}{N}\sum_{k=0}^{N-1} f(k\frac{T}{N})\frac{\sin(\pi (t-k\frac{T}{N})/\frac{T}{N})}{\sin(\pi
    (t-k\frac{T}{N})/T)}
\end{equation}

\begin{figure}[ht]
    \centering
    \begin{minipage}{.23\textwidth}
        \centering
        \includegraphics[width=.95\linewidth]{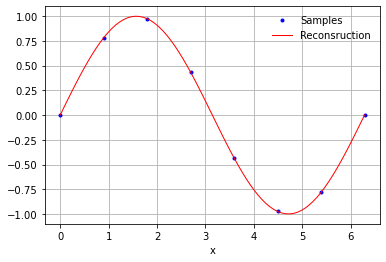}
    \end{minipage}
    \begin{minipage}{.23\textwidth}
        \centering
        \includegraphics[width=.95\linewidth]{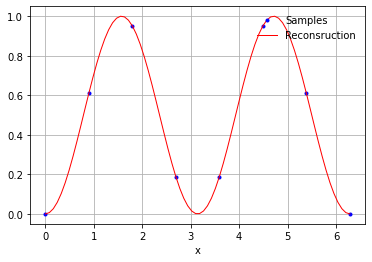}
    \end{minipage}
    \captionof{figure}{Left: Reconstruction of $f(x) =\sin(x)$ from 8 samples. Right: Reconstruction of $f(x) =\sin^2
    (x)$ from 8 samples.}
    \label{fig:exact_reconstruction}
\end{figure}

An example of exact reconstruction for functions $\sin$ and  $\sin^2$ from 8 samples on the interval $\left[ 0, 2\pi
\right]$ is illustrated in Fig.~\ref{fig:exact_reconstruction}.

We can see that an efficient quantum implementation of the reconstruction formula in
Equation~\ref{eqn:classical_interpolation} allows the exact calculation of band-limited functions at any intermediate
point of a digitally encoded interval.

\begin{figure}[ht]
    \centering
    \begin{minipage}{.23\textwidth}
        \centering
        \includegraphics[width=.95\linewidth]{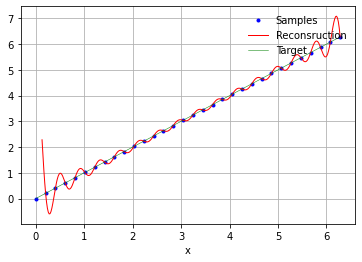}
    \end{minipage}
    \begin{minipage}{.23\textwidth}
        \centering
        \includegraphics[width=.95\linewidth]{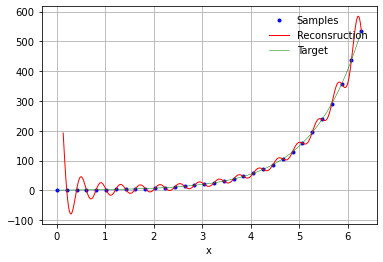}
    \end{minipage}
    \captionof{figure}{Left: Reconstruction of $f(x) = x$ from 32 samples. Right: Reconstruction of $f(x) = e^x$
        from 32 samples.}
    \label{fig:approx_reconstruction}
\end{figure}

If a function is not band-limited, it still can be interpolated at the sampling points and approximated with error
between the points as illustrated for the linear function $f(x)=x$ and the exponential function $f(x)=e^x$ in
Fig.~\ref{fig:approx_reconstruction}. These functions have infinite Fourier expansions and are not periodic.

\begin{figure}[ht]
    \centering
    \begin{minipage}{.23\textwidth}
        \centering
        \includegraphics[width=.95\linewidth]{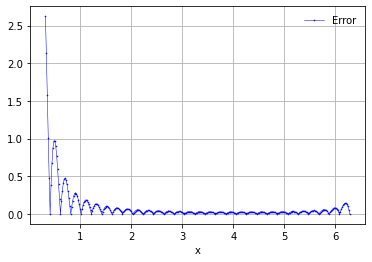}
    \end{minipage}
    \begin{minipage}{.23\textwidth}
        \centering
        \includegraphics[width=.95\linewidth]{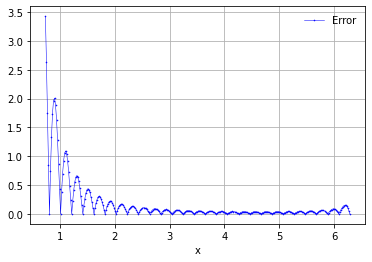}
    \end{minipage}
    \captionof{figure}{Left: Approximation error for $f(x) = x$ from 32 samples. Right: Approximation error for $f
    (x) = e^x$ from 32 samples.}
    \label{fig:approx_error}
\end{figure}

The reconstruction formula accounts only for $\frac {N-1}{2}$ of the lower frequencies of these functions. The rest
of the frequencies create the approximation error. It is also worth mentioning that the approximation error
decreases in the middle part of the sampling interval as presented in Fig.~\ref{fig:approx_error}. This behavior is
related to the non-periodicity of the function.

\subsection{\label{subsec:num_encoding}Number Encoding}

This subsection is based on ~\cite{gilliam2021foundational, CPBO}.

\begin{figure}[ht]
    \centering
    \mbox{
        \Qcircuit @C=1em @R=0em @!R {
            0 & {} & {} & {} & \gate{P(2^{m - 1}\theta)} & \qw & \qw  \\
            \vdots & & & & \ldots \\
            m-1-i & {} & {} & {} & \gate{P(2^{i}\theta)} & \qw & \qw  \\
            \vdots  & & & & \ldots \\
            m - 1 & {} & {} & {} & \gate{P(\theta)} & \qw & \qw   \\
        }
    }
    \caption{The quantum circuit for the operator $U_{\gamma} (\theta)$ applied to a quantum register with
        $m$ qubits. The circuit performs a series of Phase gates, denoted by $P$, using multiples of a given
        angle $\theta \in [-\pi, \pi)$.}
    \label{fig:geom_circuit}
\end{figure}
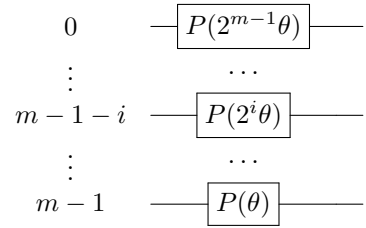

Given a quantum register with $m>0$ qubits, and an angle $\theta \in [-\pi, \pi)$, we can encode the state

\begin{equation}
    \label{eqn:geom_state_eqn}
    \begin{split}
        \ket{\gamma_{\theta}} & = \frac{1}{\sqrt{M}}\sum_{k=0}^{M-1} e^{i k\theta} \ket{k}_m \\
        & = \frac{1}{\sqrt{M}}\sum_{k=0}^{M-1}\left(\cos(k\theta) + i\sin(k\theta) \right) \ket{k}_m,
    \end{split}
\end{equation}

where $M = 2^m$, applying the unitary operator $U_{\gamma} (\theta)$
described in Fig.~\ref{fig:geom_circuit}, to an $m$-qubit register in equal superposition.

We will use the notation $V_M = [0, M)$ when we want to represent non-negative values, or $V_M = [-\frac{M}{2},
\frac{M}{2})$ if we want to represent negative values with Two's Complement.

For an integer $M$ and a real number $t \in V_M$, we define the function $c_{M,t} : \{0,\mathellipsis,M\}
\rightarrow \mathbb{R}$ as

\begin{equation*}
    c_{M,t}(k) =
    \begin{cases}
        1, & \text{if } t \text{ is an integer and } k = t \\
        0, & \text{if } t \text{ is an integer and } k \neq t \\
        \frac{1}{M} \frac{\sin(\pi (t - k))}{\sin(\frac{\pi}{M}(t-k))}, & \text{otherwise} \\
    \end{cases} \\
\end{equation*}

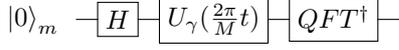
\begin{figure}[ht]
    \centering
    \mbox{
        \Qcircuit @C=0.8em @R=1em {
            & \lstick{\ket{0}_m{}} & \gate{H}   & \gate{U_{\gamma}(\frac{2\pi}{M}t)}  & \gate{QFT^\dag} & \qw \\
        }
    }
    \caption{A quantum circuit which encodes of a value $t \in V_M$ in a quantum register with $m$ qubits.}
    \label{fig:value_encoding_circuit}
\end{figure}

We can encode a number $t \in V_M$ by preparing the state $\ket{\gamma_{\frac{2 \pi}{M} t}}$,
and then applying the inverse Fourier transform ($QFT^\dagger$) as shown in Fig.~\ref{fig:value_encoding_circuit}, 
arriving at the state

\begin{equation}
    \ket{\phi_{m,t}} = \sum_{k = 0}^{M-1} e^{i \pi \frac{M -1}{M} (t - k)} c_{M,t}(k) \ket{k}_m.
    \label{eqn:fejer_state}
\end{equation}

\begin{figure}[ht]
    \centering
    \includegraphics[width=.25\textwidth]{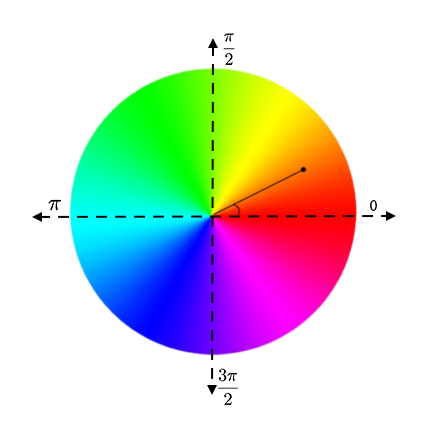}
    \caption{The phase of a complex number determines the hue on the color wheel.}
    \label{fig:color_wheel}
\end{figure}

Throughout this paper, we include bar graph visualizations of quantum states using the visualization approach
introduced in~\cite{gilliam2021foundational} where the colors correspond to the phases of the amplitudes (complex
numbers) on the color wheel in Fig.~\ref{fig:color_wheel}. Using this visualization approach, amplitudes with
positive real values (a phase of 0) will appear red and those with negative real values (a phase of $\pi$) will
appear blue.

\begin{figure*}[ht]
    \centering
    \begin{minipage}{.3\textwidth}
        \centering
        \includegraphics[width=.8\linewidth]{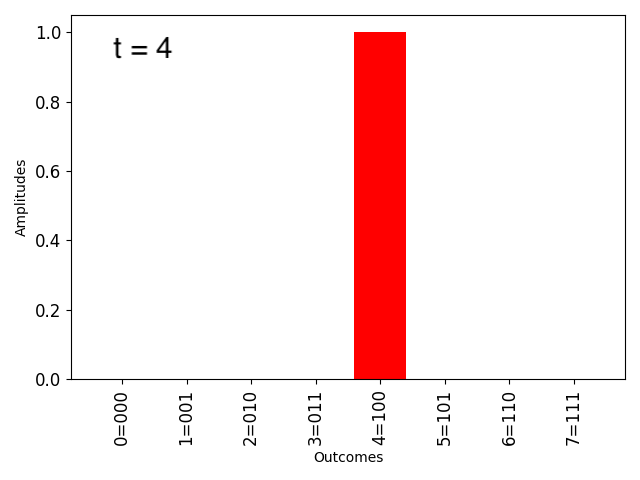}
    \end{minipage}
    \begin{minipage}{.3\textwidth}
        \centering
        \includegraphics[width=.8\linewidth]{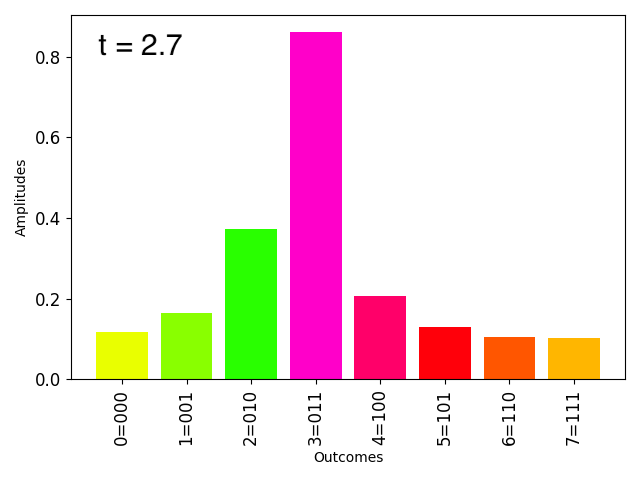}
    \end{minipage}
    \begin{minipage}{.3\textwidth}
        \centering
        \includegraphics[width=.8\linewidth]{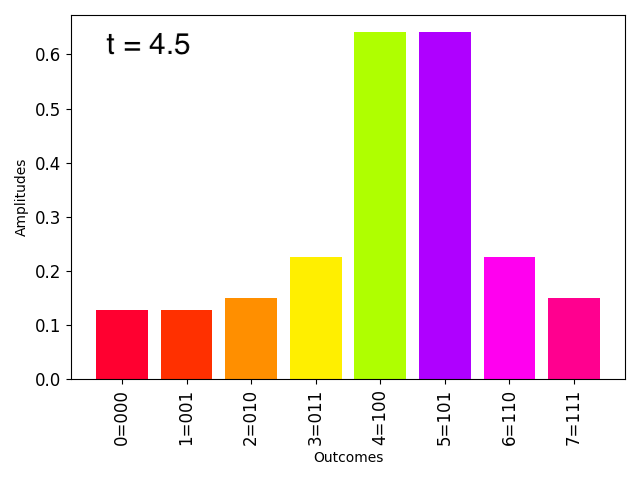}
    \end{minipage}
    \captionof{figure}{Left: Visualization of the amplitudes of the state $\ket{\phi_{3, 4}}$. The only possible outcome
    is the integer 4. Middle: Visualization of the amplitudes of the state $\ket{\phi_{3, 2.7}}$. The amplitudes
    with the two largest absolute values correspond with the integer values 2 and 3, the closest to 2.7. Right:
    Visualization of the amplitudes of the state $\ket{\phi_{3, 4.5}}$. The amplitudes with the two largest absolute
    values are equally split between the integer values 4 and 5.}
    \label{fig:encoding_examples}
\end{figure*}

Fig.~\ref{fig:encoding_examples} shows three examples of real number encoding.

If $t$ is an integer, the resulting state is $\ket{t}$ if $t \ge 0$ or $\ket{t+M}$ if $t < 0$,
encoding the Two's Complement representation of $t$, as described in ~\cite{gilliam2021foundational, CPBO}.

If $t$ is not an integer, the outcomes corresponding to the two closest integers to $t$ have the highest probabilities.
The probability of measuring one of them is at least 81\% ~\cite{nielsen_chuang_2010}. We will call the probability
distribution described above and visualized in the examples in Fig.~\ref{fig:encoding_examples} a Fej\'er
distribution.

\subsection{\label{subsec:encoding_func_realt}Encoding Real-Valued Discrete Functions With Quantum Multi-Value
Dictionaries}

In this section we describe the encoding of a polynomial of binary variables and real coefficients as an extension of
the Quantum Dictionary pattern~\cite{gilliam2021foundational}.

Let us consider two quantum registers, a key register $\ket{k}_n$ with $n$ qubits and value register $\ket{v}_m$ with
$m$ qubits, and $N=2^n, M=2^m$ computational states, respectively.

Any function $f: \{0, \mathellipsis, N-1\} \rightarrow V_M$ can be represented as a
polynomial of binary variables with real coefficients $p:\{0, 1\}^n \rightarrow V_M$
~\cite{booleanfunctions, GenInnerProduct} which can be expressed as a sum of
monomials:

\begin{equation}
    \label{eqn:sum_monomials}
    p(x_0, \mathellipsis, x_{n-1}) = \sum_{J \subseteq \{ 0, \mathellipsis, n-1 \}} c_J \prod_{j \in J}x_j,
\end{equation}

where $x_j \in \{0,1\}$ and $c_J \in \mathbb{R}$ for any $J \subseteq \{ 0, \mathellipsis, n-1 \}$.

We encode key-value pairs $(k, p(k))$, for integers $0 \le k < N$ and real values $p(k) \in V_M$, as follows:

We start by putting both registers in equal superposition:

\begin{equation*}
    \frac{1}{\sqrt{N}} \sum_{k = 0}^{N - 1} \ket{k}_n \frac{1}{\sqrt{M}} \sum_{v = 0}^{M - 1} \ket{v}_m.
\end{equation*}

For each monomial $c_J \prod_{j \in J}x_j$, we encode the real value $c_J$ using the value encoding method described
in Section~\ref{subsec:num_encoding}, controlled on the qubits in the key register corresponding to $J$. This creates
the state

\begin{equation*}
    \frac{1}{\sqrt{N}} \sum_{k = 0}^{N - 1} \ket{k}_n \ket{\gamma_{\frac{2 \pi}{M} f(k)}}_m.
\end{equation*}

where $\ket{\gamma_\theta}$ for $\theta = \frac{2 \pi}{M} f(k)$ is the state defined in Equation~\ref{eqn:geom_state_eqn}.

Then, we apply the inverse Fourier transform, creating the state

\begin{equation}
    \label{eqn:dict_state_end}
    \frac{1}{\sqrt{N}} \sum_{k = 0}^{N - 1} \ket{k}_n \ket{\phi_{m,f(k)}}_m.
\end{equation}

where $\ket{\phi_{m,t}}$ is the state defined in Equation~\ref{eqn:fejer_state}.

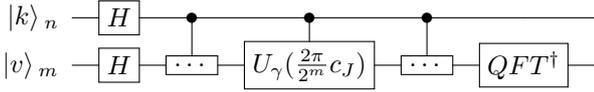
\begin{figure}[ht]
    \centering
    \mbox{
        \Qcircuit @C=1em @R=0em @!R {
            & \lstick{\ket{k}{}_n} & \gate{H} & \ctrl{1} & \ctrl{1} & \ctrl{1} & \qw & \qw  \\
            & \lstick{\ket{v}{}_m} & \gate{H} & \gate{\text{\ldots}} &  \gate{U_\gamma(\frac{2\pi}{2^{m}}c_J)} & \gate{\text{\ldots}} & \gate{QFT^\dag} & \qw
        }
    }
    \caption{The circuit for the operator $F$, applied to the key register $\ket{k}_n$ and the value register
        $\ket{v}_m$. Starting in a state of equal superposition, the operator employs several applications of the
        unitary operator $U_\gamma$, as described in Section~\ref{subsec:num_encoding}, controlled on the
        corresponding subset of key qubits $J \subseteq \{0, \ldots, n - 1\}$, whose angle parameter corresponds to
        a non-zero coefficient $c_J$. The single application of the inverse Quantum Fourier Transform ($QFT^{\dagger}$)
        at the end of the circuit decodes the periodic signal encoded by $U_\gamma$, resulting in the
        desired superposition of key-value pairs.}
    \label{fig:operator_f}
\end{figure}

We denote by $F$ the composite unitary operator that applies the sequence of gates described above to encode function
$f$ using a Quantum Multi-Valued Dictionary. The quantum circuit that represents the operator $F$ is shown in
Fig.~\ref{fig:operator_f}.

\begin{figure}[ht]
    \centering
    \includegraphics[width=.4\textwidth]{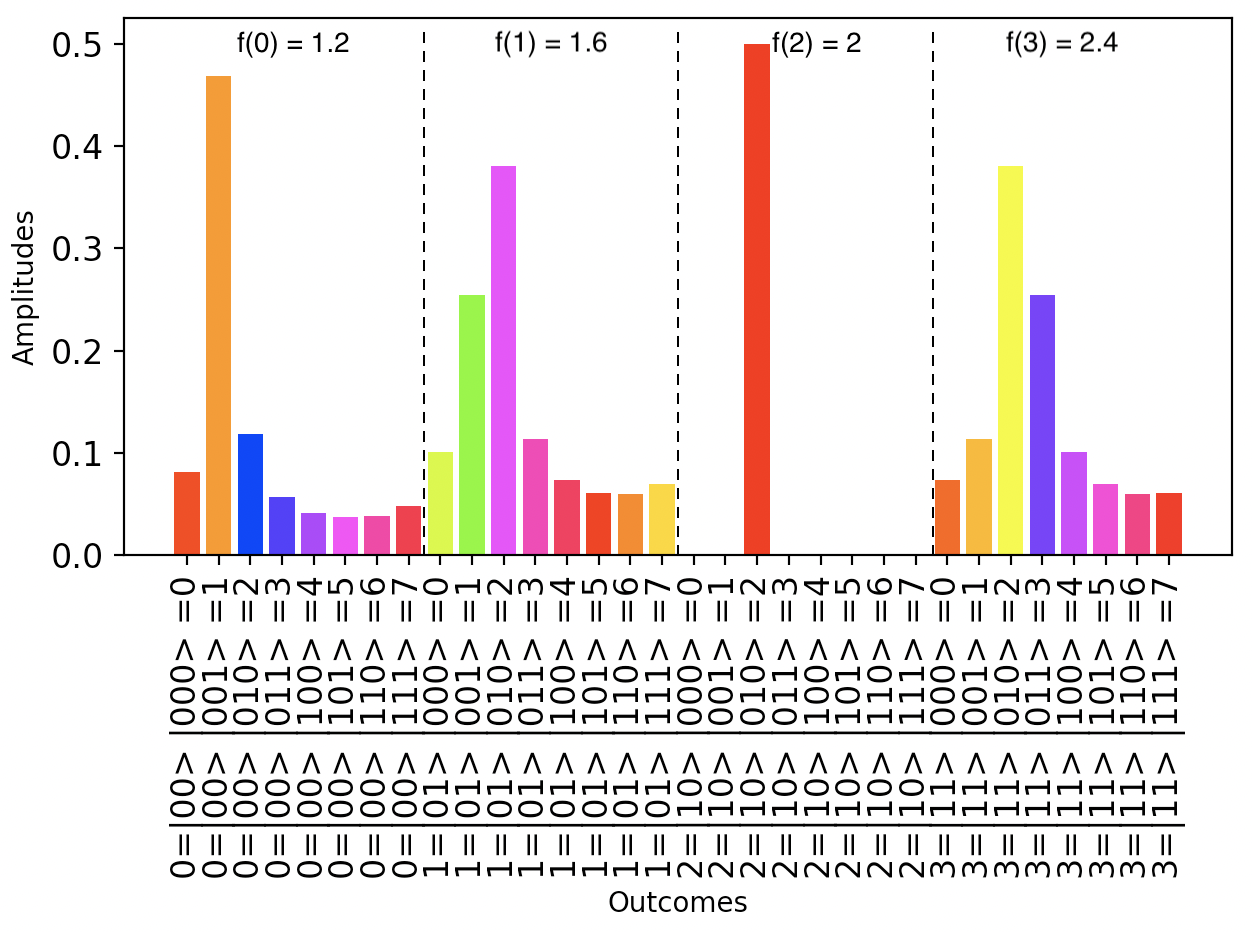}
    \caption{Visualization of the amplitudes of a quantum system after applying the operator $F$ to encode the
    function $f(k) = 1.2 + 0.4k$, using 2 qubits for the key register and 3 qubits for the value register. For each
    input $k$, where $0 \le k < 2^n$, the corresponding key-value pairs show the inputs and outputs of the function
        $f$.}
    \label{fig:poly_encoding_example}
\end{figure}

When the resulting state represented in Equation~\ref{eqn:dict_state_end} is measured, the key qubits represent the
inputs of the polynomial $p$, and the corresponding value qubits represent the outputs.
Each possible input has the same probability of being measured. An example of this encoding
method is shown in Fig.~\ref{fig:poly_encoding_example}.

\section{\label{sec:methods}Quantum Interpolation Methods}

\subsection{\label{subsec:phase_correction}Phase Corrected Number Encoding}

\begin{figure*}[ht]
    \centering
    \begin{minipage}{.3\textwidth}
        \centering
        \includegraphics[width=.8\linewidth]{encode_4}
    \end{minipage}
    \begin{minipage}{.3\textwidth}
        \centering
        \includegraphics[width=.8\linewidth]{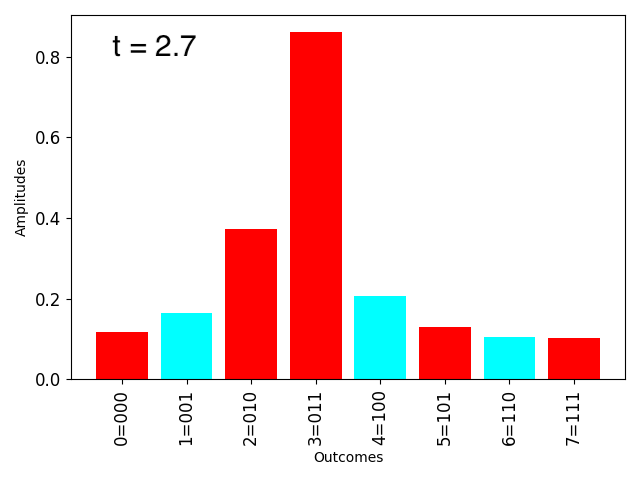}
    \end{minipage}
    \begin{minipage}{.3\textwidth}
        \centering
        \includegraphics[width=.8\linewidth]{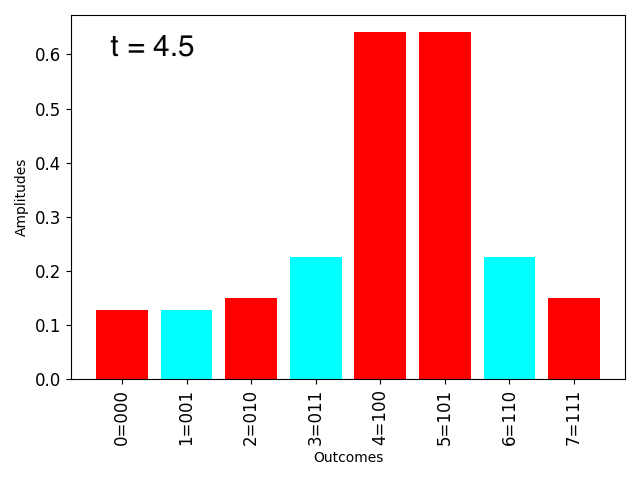}
    \end{minipage}
    \captionof{figure}{Left: Visualization of the amplitudes of the state $\ket{\iota_{3,4}}$. The resulting state has a
    probability of 1 for the outcome correlating to the two's complement of 4, which is the same as the same as the
    state $\ket{\phi_{3,4}}$ because the value encoded in the state is an integer. Middle: Visualization of the
    amplitudes of the state $\ket{\iota_{3,2.7}}$. Right: Visualization of the amplitudes of the state
        $\ket{\iota_{3,4.5}}$.}
    \label{fig:encoding_examples_real}
\end{figure*}

For a positive integer $m$ and a real number $t \in V_M$, the amplitudes of the state
$\ket{\phi_{m,t}}$ defined in Equation~\ref{eqn:fejer_state} after phase correction are the coefficients in the
interpolation theorem discussed in Equation~\ref{eqn:classical_interpolation}.

We denote the phase corrected quantum state that has real amplitudes by $\ket{\iota_{m,t}}$:

\begin{equation}
    \ket{\iota_{m,t}} = \sum_{k = 0}^{M-1} c_{M,t}(k) \ket{k}_m.
    \label{eqn:fejer_real_state}
\end{equation}

where

\begin{equation*}
    c_{M,t}(k) =
    \begin{cases}
        1, & \text{if } t \text{ is an integer and } k = t \\
        0, & \text{if } t \text{ is an integer and } k \neq t \\
        \frac{1}{M} \frac{\sin(\pi (t - k))}{\sin(\frac{\pi}{M}(t-k))}, & \text{otherwise} \\
    \end{cases} \\
\end{equation*}

The outcome probability distribution of the state is preserved (i.e. it is still a Fej\'er distribution), but the
amplitudes of the state are real numbers.

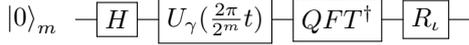
\begin{figure}[ht]
    \centering
    \mbox{
        \Qcircuit @C=0.8em @R=1em {
            & \lstick{\ket{0}_m{}} & \gate{H}   & \gate{U_{\gamma}(\frac{2\pi}{2^m}t)}  & \gate{QFT^\dag} & \gate{R_\iota} &\qw \\
        }
    }
    \caption{The encoding of a real value $t \in V_M$ using a quantum register with $m$ qubits, resulting in a state
    with real amplitudes, as defined in Eq.~\ref{eqn:fejer_real_state}.}
    \label{fig:operator_V_circuit}
\end{figure}

To prepare the phase corrected quantum state $\ket{\iota_{m,t}}$, as defined in Equation~\ref{eqn:fejer_real_state},
we apply the circuit shown in Fig.~\ref{fig:operator_V_circuit}, which includes the operator $R_\iota$ that corrects
the phase rotations.

We denote by $D_{m, t}$ the composite unitary operator shown in Fig.~\ref{fig:operator_V_circuit} that encodes a given
value $t$ by creating the state $\ket{\iota_{m, t}}$.

Fig.~\ref{fig:encoding_examples_real} shows the same three examples of number encoding as in
Fig.~\ref{fig:encoding_examples} with the phase rotations removed. Note that the amplitudes of the resulting states
appear red and blue because the amplitudes are real values.

\subsection{\label{subsec:quantum_interpolation}Quantum Amplitude Interpolation}


Considering the method for phase corrected real number encoding discussed in the previous section, as well as the
classical interpolation theorem discussed in Section~\ref{subsec:classical_interpolation}, we introduce the following
method for quantum amplitude interpolation:

\begin{theorem-non1}
    \label{quantum_interp}
    Given an integer $m > 0$ and $M = 2^m$, a function $f: \{0, \mathellipsis, M-1\} \rightarrow [0,1]$ and a
    real value $t \in V_M$ (where $V_M$ is $[0, M)$ when encoding only non-negative values, and $[-\frac{M}{2},
    \frac{M}{2})$ when encoding negative values), if a discretized version of $f$ is encoded as the amplitudes of a
    quantum state, we want an amplitude representation of the value $f(t)$.
\end{theorem-non1}

\begin{proof}[Solution]
    Assume $A$ is an operator that encodes the function $f$ into a quantum register with $m$ qubits. The operator
    $D_{m, t}$, as shown in Fig.~\ref{fig:operator_V_circuit}, prepares the state $\ket{\iota_{m,t}}$.

    We can compute the inner product of the state $\ket{\iota_{m,t}}$ and the one encoding the values of $f$ as

    \begin{equation}
        \label{eqn:fejer_and_f}
        \bra{0}_{m}A^\dagger D_{m, t}  \ket{0}_{m}  = \sum_{k = 0}^{M-1} f(k) c_{M,t}(k)
    \end{equation}

    Considering Equation~\ref{eqn:classical_interpolation}, $f(t)$ is approximated by the amplitude of $\ket{0}_{m}$
    in the state $\bra{0}_{m} A^\dagger D_{m, t}  \ket{0}_{m}$.

\end{proof}

\subsection{\label{subsec:phase_corrected_func_encoding}Phase Corrected Encoding of a Real-Valued Discrete Function}

In this section, we use the method for encoding real-valued discrete functions described in
Section~\ref{subsec:encoding_func_realt}, followed by a phase correction operator, as introduced in
Section~\ref{subsec:phase_correction}, to encode a real-valued discrete functions into a quantum state with real
amplitudes.

Assume we have two quantum registers, a key register $\ket{k}_n$ with $n$ qubits and a value register $\ket{v}_m$ with
$m$ qubits, and $N = 2^n$, $M = 2^m$. Given a function $f: \{0, \mathellipsis, N-1\} \rightarrow V_M$ expressed as a
polynomial of binary variables $p : \{0,1\}^n \rightarrow V_M$ with real coefficients, we encode key-value pairs $
(k, p(k))$, for integers $0 \le k < N$ and real values $p(k) \in V_M$, resulting in the state

\begin{equation}
    \label{eqn:dict_state_end_phase_correct}
    \frac{1}{\sqrt{N}} \sum_{k = 0}^{N - 1} \ket{k}_n \ket{\iota_{m,f(k)}}_m
\end{equation}

where $\ket{\iota_{m,t}}$ is the state defined in Equation~\ref{eqn:fejer_real_state}.

We denote by $F^\prime$ the unitary operator that encodes the phase corrected encoding of the function $f$
from Fig.~\ref{fig:operator_f_prime}.

\begin{figure*}[ht]
    \centering
    \mbox{
        \Qcircuit @C=1em @R=0em @!R {
            & \lstick{\ket{k}{}_n} & \gate{H} & \ctrl{1} & \ctrl{1} & \ctrl{1} & \qw  & \multigate{1}{R_\iota} &\qw \\
            & \lstick{\ket{v}{}_m} & \gate{H} & \gate{\text{\ldots}} &  \gate{U_\gamma(\frac{2\pi}{2^{m}}c_J)} &
            \gate{\text{\ldots}} & \gate{QFT^\dag} & \ghost{R_\iota} & \qw \\
        }
    }
    \caption{The circuit for the operator $F^\prime$, applied to the key register $\ket{k}_n$ and the value register
        $\ket{v}_m$. Starting in a state of equal superposition, the operator employs several applications of the
        unitary operator $U_\gamma$, as described in Section~\ref{subsec:num_encoding}, controlled on the
        corresponding subset of key qubits $J \subseteq \{0, \ldots, n - 1\}$, whose angle parameter corresponds to
        a non-zero coefficient $c_J$. The single application of the inverse Quantum Fourier Transform ($QFT^{\dagger}$)
        at the end of the circuit decodes the periodic signal encoded by $U_\gamma$, resulting in the
        desired superposition of key-value pairs. The operator $R_\iota$ removes the phase rotations, resulting in
        state with real amplitudes.}
    \label{fig:operator_f_prime}
\end{figure*}

\begin{figure}[ht]
    \centering
    \includegraphics[width=.4\textwidth]{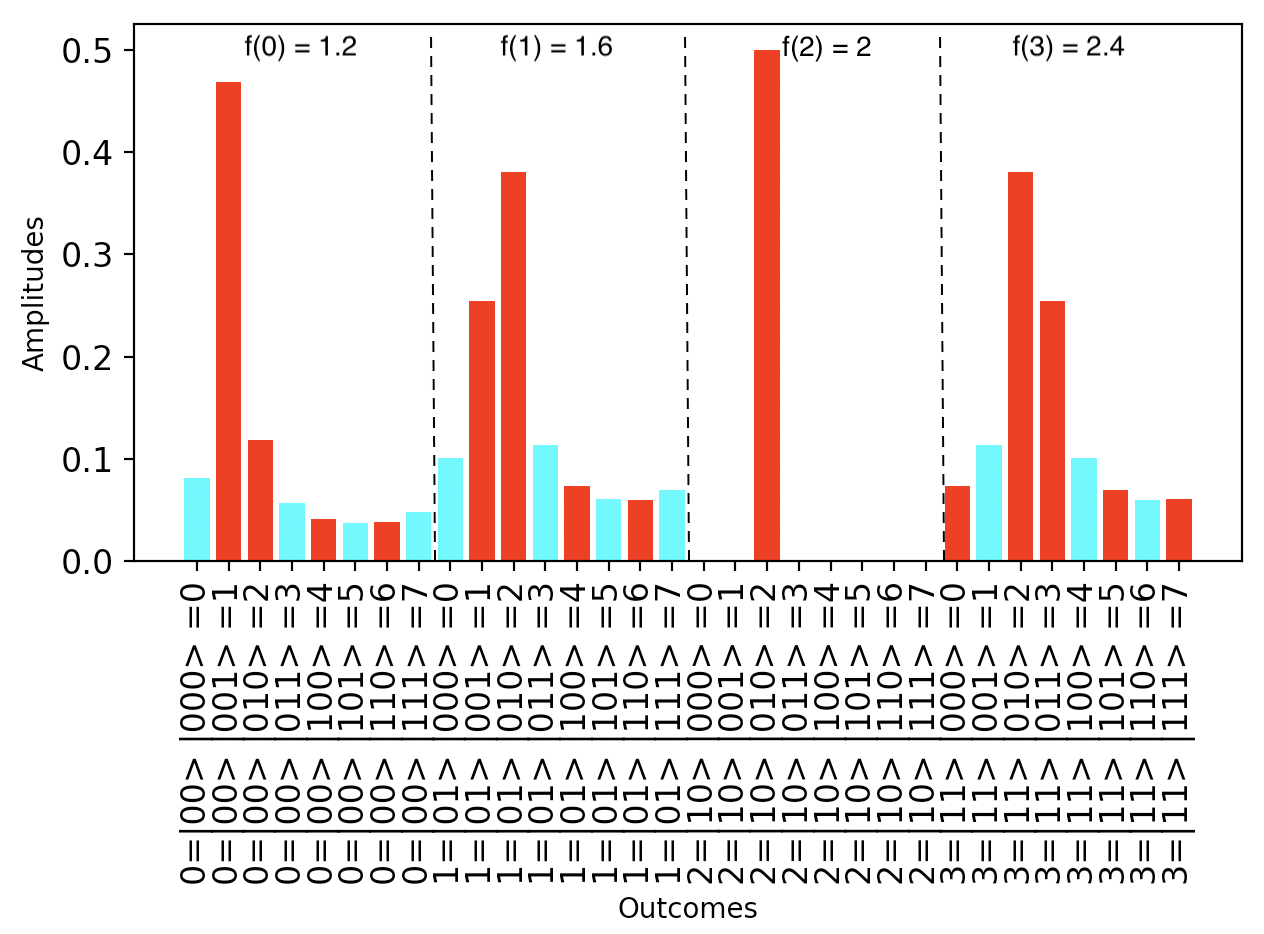}
    \caption{Visualization of the amplitudes of a quantum system after applying the operator $F^\prime$ to encode the
    function $f(k) = 1.2 + 0.4k$, using 2 qubits for the key register and 3 qubits for the value register. For each
    input $k$, where $0 \le k < 2^n$, the corresponding key-value pairs show the encoding of $f(k)$ with real
    amplitudes.}
    \label{fig:poly_encoding_example_real}
\end{figure}

Fig.~\ref{fig:poly_encoding_example_real} shows the same example in Fig.~\ref{fig:poly_encoding_example}, however,
the resulting amplitudes in Fig.~\ref{fig:poly_encoding_example_real} appear in red and blue as they are real
amplitudes.

\subsection{\label{subsec:interpolation_dict}Generalized Inner Product With Quantum Amplitude Interpolation}

We can use the quantum interpolation method to extend the generalized inner product method introduced in
~\cite{GenInnerProduct} to real-valued discrete functions.

Let us consider two quantum registers, a key register with $n$ qubits and a value register with $m$ qubits, and $N
= 2^n$, $M = 2^m$. As discussed in Section~\ref{subsec:encoding_func_realt}, given a function $f: \{0,
\mathellipsis, N-1\} \rightarrow V_M$ we can entanlge the two registers so the key-value pairs
correlate the inputs and outputs of the function $f$.

Let operator $A$ encode a given distribution of weights on the key register, creating the state

\begin{equation*}
    \label{eqn: unitary_a}
    A\ket{0}_{n} = \sum_{k = 0}^{N-1}a_k\ket{k}_n.
\end{equation*}

The combined state of both registers is

\begin{equation*}
    \label{eqn:state_after_a}
    (A \otimes I_m) \ket{0}_{n+m} = \sum_{k = 0}^{N-1}a_k\ket{k}_n\ket{0}_m.
\end{equation*}

The operator $F^\prime$ encodes the function $f$, as shown in Fig.~\ref{fig:operator_f_prime}, creating the state

\begin{equation}
    \label{eqn:state_after_f}
    \begin{split}
        F^\prime (A \otimes I_m)\ket{0}_{n+m} & = F^\prime\left(\sum_{k = 0}^{N-1}a_k\ket{k}_n\ket{0}_m\right) \\
        & = \sum_{k = 0}^{N-1}a_k\ket{k}_n\ket{\iota_{m, f(k)}}_m.
    \end{split}
\end{equation}

Let $B$ be an operator that encodes weights/hashes in the value register

\begin{equation*}
    \label{eqn:unitary_b}
    B\ket{0}_m = \sum_{v = 0}^{M-1}b_v\ket{v}_m.
\end{equation*}

Applying $(H^{\otimes n} \otimes B^\dagger)$ to the state in Equation~\ref{eqn:state_after_f} results in the state

\begin{equation}
    \label{eqn:generalized_inner_product}
    \begin{split}
        \bra{0}_{n+m}(H^{\otimes n} \otimes B^{\dagger}) F^\prime (A \otimes I_m)\ket{0}_{n+m} = \\
        \sum_{k = 0}^{N-1} a_k \sum_{v=0}^{M-1} b_v c_{M,f(k)}(v).
    \end{split}
\end{equation}

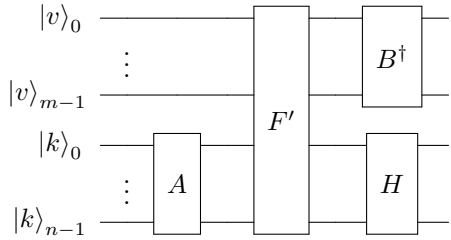
\begin{figure}[ht]
    \centering
    \mbox{
        \Qcircuit @C=1em @R=1em {
            & \lstick{\ket{v}_0{}}  & \qw & \qw & \qw & \multigate{5}{F^\prime}    & \qw   &\multigate{2}{B^{\dagger}}   &\qw \\
            & \lstick{\textcolor{white}{0}} & \vdots \\
            & \lstick{\ket{v}_{m-1}}    & \qw & \qw & \qw & \ghost{F^\prime}   & \qw   &\ghost{B^{\dagger}}    & \qw \\ 
            & \lstick{\ket{k}_0{}}  & \qw & \multigate{2}{A}    & \qw & \ghost{F^\prime}   & \qw   & \multigate{2}{H} &  \qw \\
            & \lstick{\textcolor{white}{0}} & \vdots \\
            & \lstick{\ket{k}_{n-1}}    & \qw & \ghost{A}   & \qw & \ghost{F^\prime}   & \qw   & \ghost{H} & \qw   \\ 
        }
    }
    \caption{The quantum circuit for computing generalized inner products with interpolation, using a quantum system
    with a key register with $n$ qubits $\ket{k}_n$, and a value register with $m$ qubits $\ket{v}_m$. The operator
        $A$ encodes a given distribution of weights on $\ket{k}_n$, and the operator $F^\prime$ encoding the inputs and
    outputs of a real-valued discrete function $f$ using the phase corrected real value encoding method, followed by
        the operator $B^\dagger$ on $\ket{v}_m$ and $H^{\otimes n}$ on $\ket{k}_n$. Applying the circuit results in
        the desired inner product as
    the amplitude $\ket{0}_{n+m}$.}
    \label{fig:dict_circuit}
\end{figure}

Let us consider the following problem context and solution in which the quantum interpolation method allows for the
generalized inner product method ~\cite{GenInnerProduct} to be applied to real-valued discrete functions:

\begin{theorem-non2}
    \label{cont_weighted_sum}
    Given integers $n, m > 0$ and $N = 2^n, M = 2^m$, weights $w_k \in \mathbb{R}$ defined for integers $0 \le k < N$,
    a weight/hash function $h: V_M \rightarrow \mathbb{R}$ and a function $f: \{0, \mathellipsis, N-1\} \rightarrow
    V_M$ (where $V_M$ is $[0, M)$ when encoding only non-negative values, and $[-\frac{M}{2}, \frac{M}{2})$ when
    encoding negative values), we are interested in calculating the weighted sum of weighted/hashed function values,

    \begin{equation*}
        \sum_{k=0}^{N-1}w_k h(f(k)).
    \end{equation*}
\end{theorem-non2}

\begin{proof}[Solution]
    Let $A$ be an operator that prepares a state $\sum_{k =
    0}^{N-1}a_k\ket{k}_n$, with $a_k = a w_k$, for $0 \le k < N$, where $a \in \mathbb{R}$ is a common factor, and
    let $B$ be an operator that prepares the state $\sum_{v = 0}^{M-1}b_v\ket{v}_m$ with $b_v = b h(v)$, for $0 \le v
    < M$, where $b \in \mathbb{R}$ is a common factor, and $F^\prime$ be the operator that encodes the real-value
    discrete function $f$.

    Then:

    \begin{equation*}
        \begin{split}
            E_{\ket{0}} & \coloneqq \bra{0}_{n+m} (H^{\otimes n} \otimes B^{\dagger}) F^\prime (A \otimes I_m)
            \ket{0}_{n+m} \\
            & = \sum_{k=0}^{N-1} w_k \sum_{v=0}^{M-1} h_v c_{M, f(k)}(v) \\
            & \approx \frac{ab}{\sqrt{N}} \sum_{k=0}^{N-1}w_k h(f(k)).
        \end{split}
    \end{equation*}

    Therefore,

    \begin{equation*}
        \sum_{k=0}^{N-1}w_k h(f(k)) \approx \frac{\sqrt{N}}{ab} E_{\ket{0}}
    \end{equation*}

    with $E_{\ket{0}} = \bra{0}_{n+m} (H^{\otimes n} \otimes B^{\dagger}) F^\prime (A \otimes I_m)\ket{0}_{n+m}$ being the
    amplitude of $\ket{0}_{n+m}$ at the end of the computation, which can be estimated using amplitude estimation
    algorithms.
\end{proof}

\begin{application1}
    If $B$ is the operator $L_m$ that encodes $h$ as the identity, i.e. $h(v) = v$ for $0 \le v < M$, and $b =
    \frac{1}{\sqrt{\sum_{v=0}^{M-1} v^2}} = \sqrt{\frac{6}{(M-1)M(2M-1)}}$ we obtain a canonical way to compute the
    expected value of a function $f : {0, \mathellipsis, N - 1} \rightarrow [0, M)$:

    \begin{equation*}
        \sum_{k=0}^{N-1}w_k f(k) = \frac{\sqrt{N}}{a} \sqrt{\frac{(M-1)M(2M-1)}{6}} E_{\ket{0}}
    \end{equation*}

    with $E_{\ket{0}} = \bra{0} (H^{\otimes n} \otimes L_m^{\dagger}) F^\prime (A\otimes I_m)\ket{0}_{n+m}$ being the
    amplitude of $\ket{0}_{n+m}$ at the end of the computation.
\end{application1}

\section{\label{sec:applications}Applications of Quantum Amplitude Interpolation}

\subsection{\label{subsec:interpolation_example_normal}Quantum Interpolation of a Normal Distribution Approximation}

In this example, we assume the availability of the state preparation operator introduced in ~\cite{GenInnerProduct},
denoted by $N_{2,m}$, which encodes an approximation of a normal distribution, resulting in the state

\begin{equation}
    \label{eqn:sin4_state}
    \begin{split}
        N_{2, m} \ket{0}_m & = \ket{\nu_2}_m\\
        & = \sqrt{\frac{8}{3M}}\sum_{k = 0}^{M-1}\sin^2\left(k\frac{\pi}{M}\right) \ket{k}_m
    \end{split}
\end{equation}

where $M = 2^m$.

As an example, given a quantum system with $m = 6$ qubits, and the function $f(k) = \sqrt{\frac{8}{3M}} \sin^2(k
\frac{\pi}{M})$ for $0 \le k < 2^m$, and the value $t = 44.8$, we encode an amplitude representation of an
approximation for $f(t)$.

Using the Quantum Interpolation pattern~\ref{quantum_interp}, where the unitary operator $D_{m, t}$ prepares the state
$\ket{\iota_{6, 44.8}}$ as defined in Equation~\ref{eqn:fejer_real_state} and the operator $A = N_{2,6}$ encodes the
function $f$ resulting in the state $\ket{\nu_2}_6$ as described above.

We can compute the inner product of the states $\ket{\iota_{6, 44.8}}$ and$\ket{\nu_2}_6$ as

\begin{equation*}
    \bra{0}_{m}A^\dagger D_{m, t}  \ket{0}_{m}  = \sum_{k = 0}^{M-1} f(k) c_{M,t}(k)
\end{equation*}

Running the quantum computation in a simulator yields $\bra{0}_{m}A^\dagger D_{m, t} \ket{0}_{m} = 0.1336$ (the
amplitude of $\ket{0}_m$). This result is equal to a classical computation of the interpolation of $t$, as defined in
Equation~\ref{eqn:classical_interpolation}, results in $0.1336$.

If we run this same computation for different values of $t \in V_M$, we obtain values because $f$ is a periodic,
band-limited function, as dicussed in Section~\ref{subsec:classical_interpolation}.
Fig.~\ref{fig:identity_interp} shows the result of the quantum interpolation computation versus the exact values of the
function.

\begin{figure}[ht]
    \centering
    \includegraphics[width=.35\textwidth]{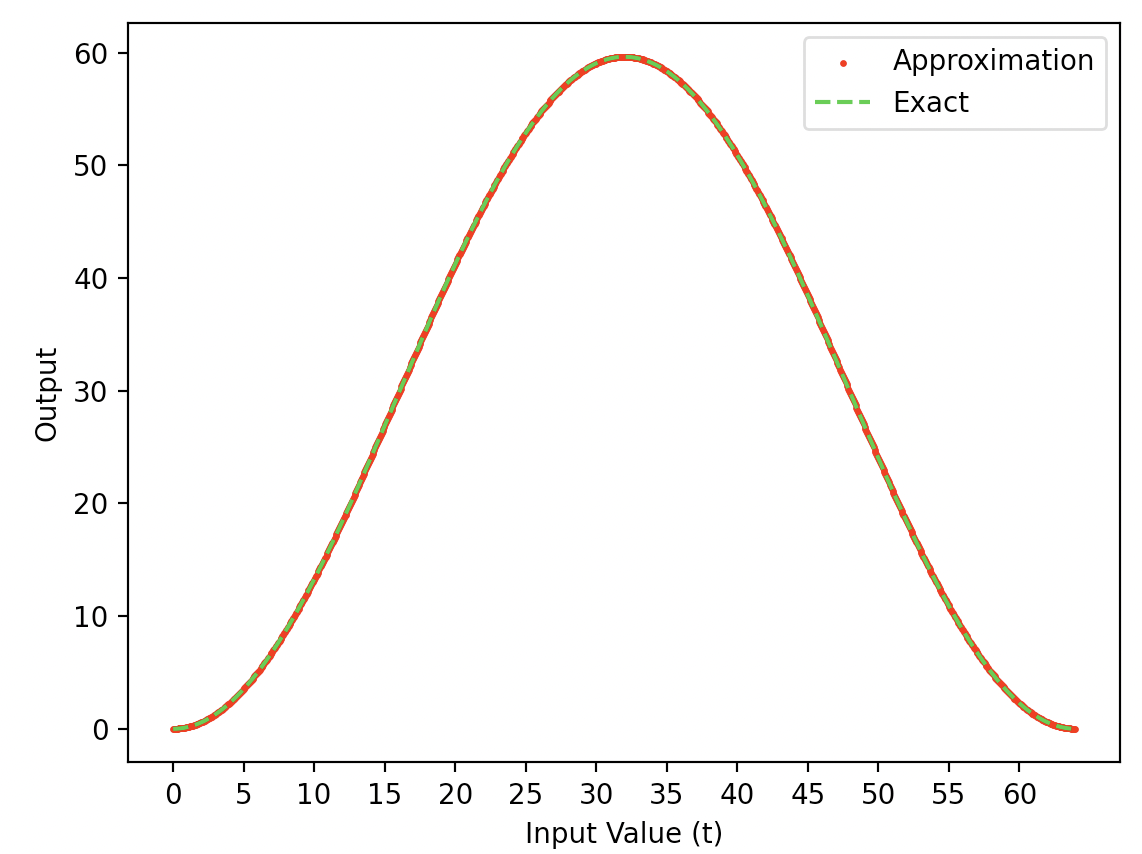}
    \caption{Exact values of the function $f(k) = \sqrt{\frac{8}{3M}} \sin^2(k \frac{\pi}{M})$ versus the amplitude
    representation of the approximation of $f$ using Quantum Interpolation. The approximations are from a simulation
    of the Quantum Interpolation method using a quantum system with $m = 6$ qubits.}
    \label{fig:sin2_interp}
\end{figure}

\subsection{\label{subsec:interpolation_example_linear}Quantum Interpolation of a Linear Function}

As an example, given a quantum system with $m = 6$ qubits, the normalized identity function $f(k) = \sqrt{\frac{6}{
    (M-1)M(2M-1)}} k$ where $M = 2^m$, for $0 \le k < M$, and a value $t = 44.8$, we encode an amplitude
representation of an approximation for $f(t)$.

In this example, we assume the availability of the state preparation operator introduced in ~\cite{GenInnerProduct},
denoted by $L_m$, which encodes an approximation of the identity function, resulting in the state

\begin{equation}
    \label{eqn:heuristic_lin}
    \begin{split}
        \ket{\lambda}_m & = L_m \ket{0}_m \\
        & = \sqrt{\frac{6}{(M-1)M(2M-1)}}\sum_{k = 0}^{M-1}k\ket{k}_m.
    \end{split}
\end{equation}

Using the Quantum Interpolation pattern~\ref{quantum_interp}, we can compute the inner product of the states
$\ket{\iota_{6,t}}$ and $\ket{\lambda}_6$ as

\begin{equation*}
    \bra{0}_{m}A^\dagger D_{m, t}  \ket{0}_{m}  = \sum_{k = 0}^{M-1} f(k) c_{M,t}(k).
\end{equation*}

Running the quantum computation in a simulator yields $\bra{0}_{m}A^\dagger D_{m, t} \ket{0}_{m} = 0.1533$ (the
amplitude $\ket{0}_m$). A classical computationn of the interpolation of $t$, as defined in
Equation~\ref{eqn:classical_interpolation}, results in $0.1546$. If we multiply $0.1533$ by the normalization factor
$\sqrt{\sum_{k = 1}^{M - 1} k^2} = 292.137$ we get the value $44.79$ which is an approximation for $f(44.8) = 44.8$.
Note that the more qubits used, the better this approximation will be.

\begin{figure}[ht]
    \centering
    \includegraphics[width=.35\textwidth]{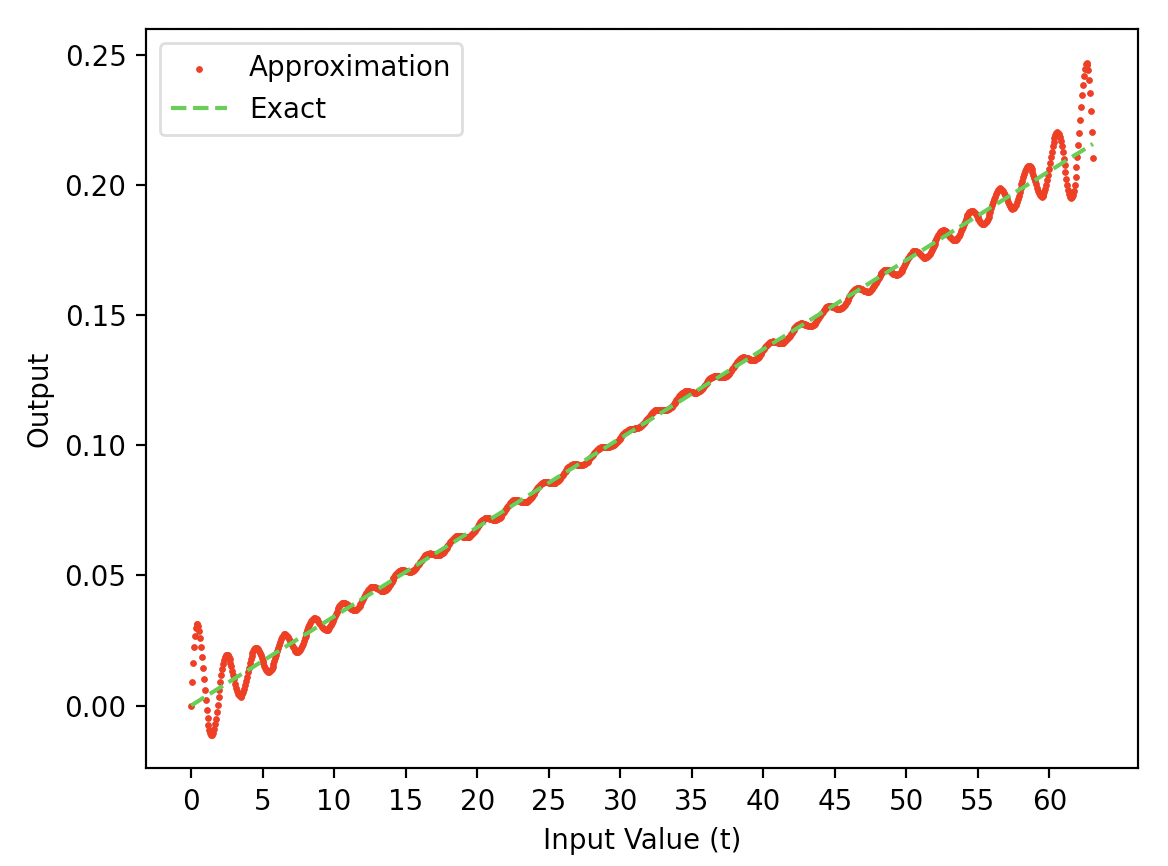}
    \caption{Exact values of the identity function $f$ versus the amplitude representation of the approximation of
        $f$ using Quantum Interpolation. The approximations are from a simulation of the Quantum Interpolation
        method using a quantum system with $m = 6$ qubits.}
    \label{fig:identity_interp}
\end{figure}

If we run this same computation for different values of $t \in V_M$, the accuracy of the interpolation varies with
the distance between $t$ and the closest integer. Fig.~\ref{fig:identity_interp} shows the varying accuracy of
approximations. Note that the function $f$ is not periodic and not band-limited, resulting in varying error
between the intepolation and $f$, as discussed in Section~\ref{subsec:classical_interpolation} and shown in
Fig.~\ref{fig:approx_reconstruction}.

\subsection{\label{subsec:expected_value_example}Expectation of Real-Valued Discrete Functions}

In this example, we assume the availability of the state preparation operator introduced in ~\cite{GenInnerProduct},
denoted by $N_{2,m}$, which encodes an approximation of a normal distribution, resulting in
the state defined in Equation~\ref{eqn:sin4_state}.

Given integers $n, m > 0$, where $N = 2^n$ and $M = 2^m$ and a function $f: \{0, \mathellipsis, N-1\} \rightarrow
V_M$, assume we want to compute the weighted sum

\begin{equation*}
    \sum_{k=0}^{N-1}w_k f(k)
\end{equation*}

for weights $w_k = \sin^2(k\frac{\pi}{N})$, where $0 \le k < N$.

As an example, given a quantum system with $n = 3$ and $m = 4$, and the function $f$, represented as the binary
polynomial

\begin{equation}
    \label{eqn:bin_poly_expected_value}
    p(k_0, k_1, k_2) = 0.725 + 2.451 k_1 + 2.716 k_0 + 1.321 k_0 k_2
\end{equation}

for $(k_0, k_1, k_3) \in \{0, 1\}^n$, with $k = \sum_{j=0}^{n-1}k_j 2^j$ being the binary expansion of $k$, as
described in Section~\ref{subsec:encoding_func_realt}.

Using the Controlled Weighted Sum~\ref{cont_weighted_sum} pattern, where $A = N_{2,3}$, and $B = L_{4}$, as
defined in Equation~\ref{eqn:heuristic_lin}, $a = \sqrt{\frac{8}{3N}}$, and $b = \sqrt{\frac{6}{(M-1)M(2M-1)}}$ we
obtain the result

\begin{equation}
    \label{eqn:expected_discrete}
    \begin{split}
        \sum_{k=0}^{N-1}w_k f(k) & = \sqrt{N} \sqrt{\frac{3N}{8}} \sqrt{\frac{(M-1)M(2M-1)}{6}} \\
        & \cdot \bra{0}_{m+m} (H^{\otimes n} \otimes B^{\dagger}) F^\prime (A \otimes I_m)\ket{0}_{n+m}
    \end{split}
\end{equation}

\begin{figure}[ht]
    \begin{tabular}{c}
        \includegraphics[width=0.9\linewidth]{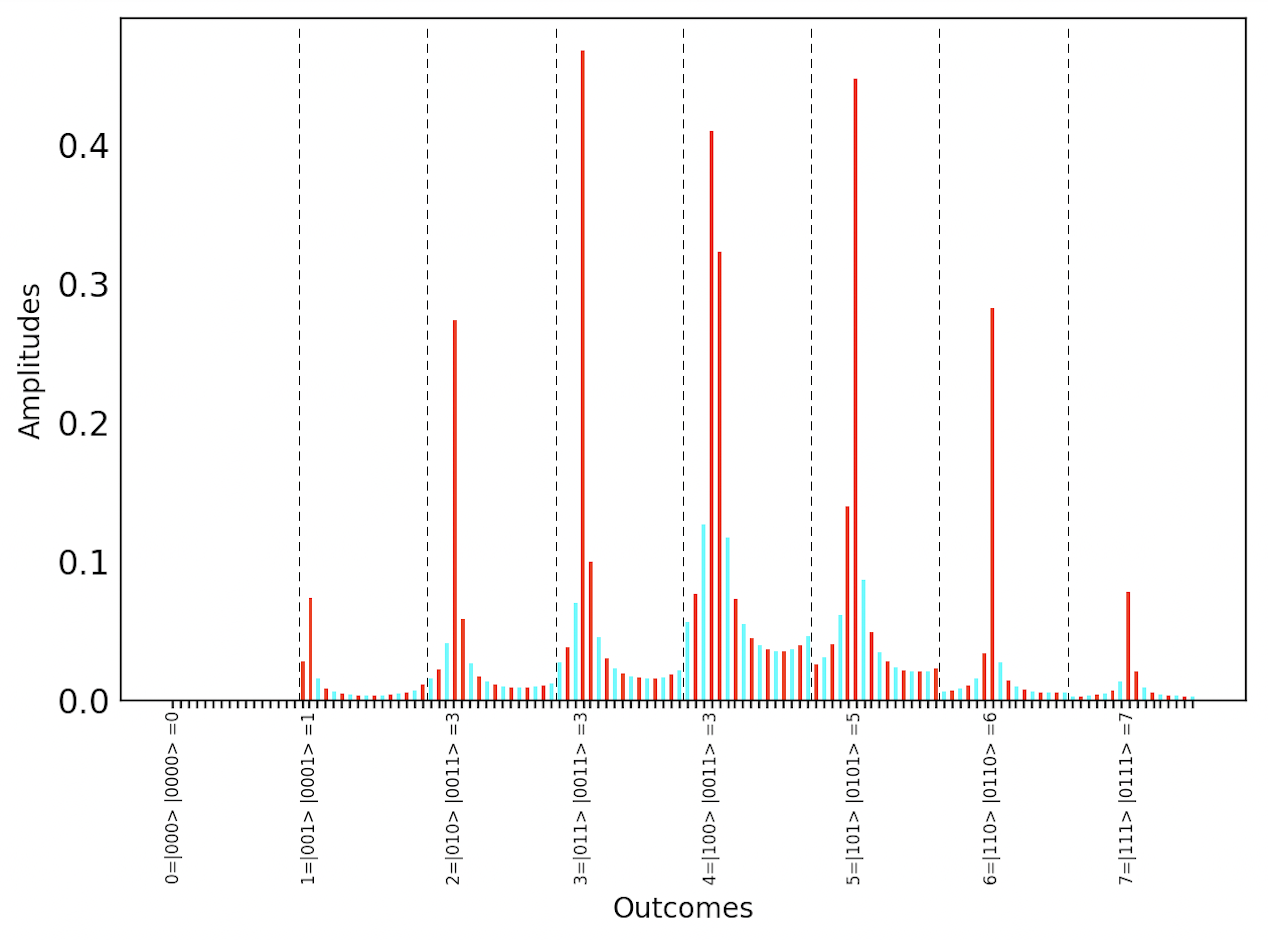} \\
        \includegraphics[width=0.9\linewidth]{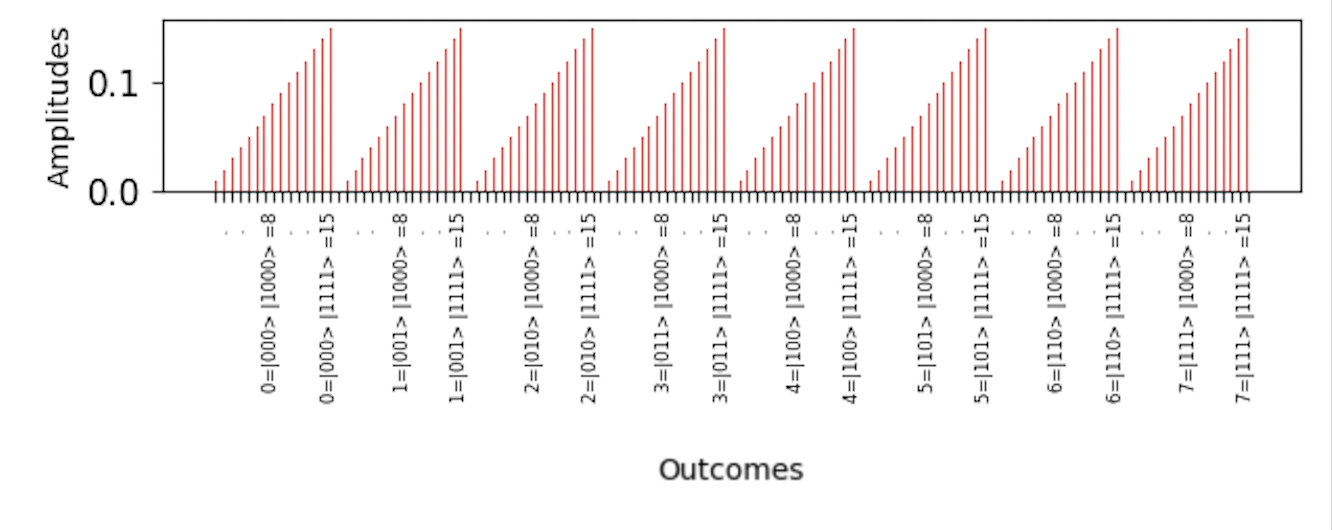} \\
    \end{tabular}
    \captionof{figure}{Top: Visualization of the amplitudes of a quantum system after applying the operator $A =
    N_{2,3}$, using 3 qubits for the key register and and 4 qubits for the value register. The key-value pairs
    on the x-axis show the result of encoding the binary polynomial in Eq.~\ref{eqn:bin_poly_expected_value}. The
    distribution of weights applied to the key register (as in Eq.~\ref{eqn:sin4_state}) is an approximation of a
    normal distribution in the amplitudes. Bottom: Visualization of amplitudes of a quantum system, using 3 qubits
    for the key register and 4 qubits for the value register, after applying the operator $B = L_4$ and Hadamard
    gates on the key register. Encoding the linear function in the value register (as in Eq.~\ref{eqn:heuristic_lin})
        results in the identity function implemented in amplitudes, repeated for each key value.}
    \label{fig:dict_normal_identity}
\end{figure}

Running the quantum computation in a simulator yields $\bra{0} (H^{\otimes n} \otimes B^{\dagger}) F^\prime (A \otimes
I_m) \ket{0}_{n+m} = 0.0879$ (the amplitude of $\ket{0}_{n+m}$), and from Equation~\ref{eqn:expected_discrete} we
obtain $\sum_{k=0}^{N-1}w_kf(k) \approx 15.1555$.  A direct classical calculation gives $\sum_{k=0}^{N-1}w_kf(k)
\approx 15.9130$.

In order to improve the precision, we add more qubits to the value register to facilitate the encoding of scaled
coefficients. If we add 6 qubits to the value register, which initially had 4 qubits, we can encode the coefficients
scaled by a factor of 64. Running the same computation as described above, where $A = N_{2,3}$, and $B = L_{10}$, $a
= \sqrt{\frac{8}{3N}}$, and $b = \sqrt{\frac{6}{(M-1)M(2M-1)}}$ we obtain the result $\bra{0} (H^{\otimes n} \otimes
B^{\dagger}) F^\prime (A \otimes I_m) \ket{0}_{n+m} = 0.0110$. From Equation~\ref{eqn:expected_discrete} we
obtain $\sum_{k=0}^{N-1}w_kf(k) \approx 15.9186$, which is a much closer approximation for the classical result
$\sum_{k=0}^{N-1}w_kf(k) \approx 15.9130$.

For a given number of available qubits, we compare the method used above with approximating the coefficients by
integers. If we are limited to 10 value qubits, as above, we can encode the largest integer coefficients that can be
accomodated by the value register. Running the same computation as described above, with the binary polynomial $p
(k_0, k_1, k_2) = 725 + 245 k_1 + 272 k_0 + 132 k_0 k_2$, we obtain $\sum_{k=0}^{N-1}w_kf(k) \approx 15.94$.
Depending on the coefficients, the error will vary.

\section{\label{sec:experiments}Experiments on Quantum Hardware}

The experiments discussed in this section were run on IBM quantum devices powered by
IBM Quantum Falcon Processors ~\cite{IBMQServices}.

\paragraph{Quantum Interpolation of a Normal Distribution Approximation.}

In this experiment we performed the expected value computation using quantum interpolation discussed in
Section~\ref{subsec:interpolation_example_normal} on the ibm{\_}perth 7-qubit device with quantum volume 32. The
average readout assignment error at the time of the experiments was 1.65\%. Each run of the experiment was performed
with 4000 shots.

The average result of the computation from 10 runs was $\bra{0}_{m}A^\dagger D_{m, t} \ket{0}_{m} = 0.1369$. In the
best experiment, the result of the computation was $\bra{0}_{m}A^\dagger D_{m, t} \ket{0}_{m} = 0.1342$ (the
amplitude of $\ket{0}_m$). A classical computation yeilds $0.1336$.

\paragraph{Quantum Interpolation of a Linear Function.}

In this experiment we performed the expected value computation discussed in
Section~\ref{subsec:interpolation_example_linear} on the ibmq{\_}guadalupe 16-qubit device with quantum volume 32.
The average readout assignment error at the time of the experiments was 1.82\%. Each run of the experiment was
performed with 4000 shots.

The average result of the 10 runs was $\bra{0}_{m}A^\dagger D_{m, t} \ket{0}_{m} = 0.1347$. In the best
experiement, the result of the computation was in $\bra{0}_{m}A^\dagger D_{m, t} \ket{0}_{m} = 0.1541$ (the
amplitude $\ket{0}_m$). A classical computation of the interpolation of $t$, as defined in
Equation~\ref{eqn:classical_interpolation}, results in $0.1546$. If we multiply $0.1541$ by the normalization factor
$\sqrt{\sum_{k = 1}^{M - 1} k^2} = 292.137$ we get the value $45.02$ which is an approximation for $f(44.8) = 44.8$.

\section{\label{sec:conclusions}Concluding Remarks and Future Directions}

The Quantum Amplitude Interpolation implementation presented in this paper allows real numbers to be digitally
encoded in quantum states and used in inner product computations. As a consequence, certain approximations can be
made more efficient and precise. For example, our approximation of normal distributions by trigonometric expressions
can be made continuous instead of only discrete.

The exact interpolation method can be used to convert outputs of other algorithms, e.g. Quantum Phase Estimation,
to an amplitude that can be used in subsequent computations.

The interpolation method can be applied to approximate other distributions or functions, e.g. linear ones, with a
manageable approximation error by shifting values to avoid extreme frequencies.

Further investigation could be conducted on the design of interpolators that better suit specific contexts.

\bigskip
\acknowledgements

The authors would like to thank Abhijit Rao for helping with the development of this manuscript.\\

The views expressed in this article are those of the authors and do not represent the views of Wells Fargo. This
article is for informational purposes only. Nothing contained in this article should be construed as investment advice.
Wells Fargo makes no express or implied warranties and expressly disclaims all legal, tax, and accounting implications
related to this article.

UC Berkeley and the UC Berkeley logo are trademarks of UC Regents.

We acknowledge the use of IBM Quantum services for this work. The views expressed are those of the authors, and do
not reflect the official policy or position of IBM or the IBM Quantum team.

\bibliographystyle{unsrtnat}
\bibliography{main}

\onecolumn\newpage
\renewcommand\floatpagefraction{0.9}
\appendix

\end{document}